# The role of solute concentration in interface instability during alloy solidification: A viewpoint from the free energy

Fengyi Yu[a], Qiaodan Hu[a,*], Jianguo Li[a]

[a] Shanghai Key Laboratory of Materials Laser Processing and Modification, School of Materials Science and Engineering, Shanghai Jiao Tong University, Shanghai 200240, PR China

[*] Corresponding author, E-mail address: qdhu@sjtu.edu.cn, Tel: 0086-21-54744246

**Abstract**

Solidification structures are determined by the interaction between the interfacial processes and transport processes of heat and solute. In this paper, we investigate planar instability in directional solidification. Firstly, the interfacial evolution at the initial growth stage is simulated, indicating the planar instability is represented by the transition from the planar to the cellular. Secondly, to represent the history-dependence of solidification, constant thermal gradient G and varying pulling speed $V_P$ are used in the simulations. The results indicate the cooling rate R ( = G*$V_P$) dominates the overall propagation speed of the interface, to maintain the local thermodynamic equilibrium. The solute segregation determines the stability of the interface, by changing the excess free energy at the interface and corresponding interface energy. Finally, the simulations of the grains with different preferred crystallographic orientations are performed, indicating the surface energy and its anisotropy do not affect the solute diffusion and planar growth. The results also verify the conclusion that solute segregation influences the interface energy and results in interface instability. On the other hand, for the planar-cellular transition, the minimum surface stiffness rule is more suitable than the maximum surface energy rule. The influence of the solute concentration on the excess free energy and interface energy can be applied to other solidification patterns induced by the interface instability, which will be studied in the future.

# 1. Introduction

Solidification structures dominate the properties of the components. The accurate prediction of solidification structures could provide theoretical basis for optimizing the parameters, for which solidification dynamics needs to be studied. Due to different characteristics of physical processes at different scales, the investigation of solidification dynamics has been a long standing challenge [1,2]. From the viewpoint of mesoscale, the solidification structures are dominated by the interaction between the interfacial processes and transport processes of heat and solute [3-5]. The diffusive nature of the transport processes, including spatial and temporal evolution, gives rise to the morphological instability of interface, resulting in different patterns of solidification structure.

The planar instability appears firstly during the evolution of solidification patterns and influences the subsequent stages significantly [6,7], which deserves systematic investigation. Chalmers et al. [8] analyzed the heat and solution balance at a moving solid/liquid (S/L) interface. They gave the idea of Constitutional Supercooling (CS) and established the CS criterion of morphological instability by expression $G*V_P \geq \Delta T_0/D_L$. In the expression, $G$ is the thermal gradient, $V_P$ is the pulling speed, $\Delta T_0$ is the solidification temperature range in the phase diagram, and $D_L$ is the solute diffusion coefficient in the liquid. Although it reveals the thermodynamic essence of interface stability, the CS theory does not account for the transport processes of heat and solute. Mullins and Sekerka (MS) [9,10] analyzed the stability of a crystal, based on a dynamic approach in which the equations governing heat flow and solute diffusion are solved simultaneously while allowing for a change of shape due to a perturbation. Compared with the CS theory, the MS theory is based on a dynamic approach, considering the interplay of diffusion transport and interface energy, reflecting the spatial and temporal evolution of solidification. However, the MS theory is performed assuming a steady-state planar interface, neglecting the time-dependence of diffusion transport. By assuming the solute concentration field evolves with time, Warren and Langer (WL) [11] extended the MS theory to non-steady-state dynamics. Their analysis indicates solidification evolution is history-dependent, depending on the detailed way in which the sample is prepared and set in motion. The interface instability predicted by the WL theory agrees well with experimental observations of real time synchrotron X-ray radiography [12], as well as the dynamics of dendritic array [13], demonstrating its validity. By combining the time-dependent linear stability analysis in the WL theory with the Fourier synthesis, Wang et al. [14] developed a simple model to predict the morphological evolution of the interface directly at the initial growth stage. The model is verified by experimental observations under steady-state conditions [15]. Subsequently, Dong et al. [16,17] modified this model from the steady-state condition to the non-steady-state condition, which can represent the time-dependent G and $V_P$. The onsets of initial instability under the same G and $V_P$, with different increasing rates of $V_P$, were simulated out. The results illustrate the increasing rate of $V_P$ does influence the solidification evolution, including the incubation time and average wavelength of the planar instability. This study demonstrates, rather than having a unique relationship with solidification conditions, the microstructure evolution depends on the detailed way the conditions were achieved. To date, considerable investigations of interface instability dynamics have been made. However, the theoretical models involve many approximations and simplifications, resulting from the constraint that the solutions of analytical and semi-analytical models can only apply under simple conditions. The analytical models can hardly handle the complex morphologies of interface and the relative effects. Moreover, the as-simplified conditions are far from the realistic solidification processes, limiting the application of these theoretical models.

Compared with the analytical method, the numerical method could solve the equations under complex conditions, having the advantage of simulating the realistic solidification processes. As a representative, the Phase-Field (PF) method combines the insights from thermodynamics and the dynamics of transport process, which has solid physical foundations [18-22]. Moreover, since it avoids the shape error caused by tracking interface during simulation, the PF method has greatly high numerical accuracy, which has been one of the most powerful numerical methods for predicting solidification [18-24]. By introducing a phenomenological "Anti-Trapping Current" (ATC) term [25], the PF model can simulate the alloy solidification quantitatively. The PF model has been applied into increasingly complex conditions, from isothermal solidification [25], directional solidification [26,27] to melt pool solidification [28-30]. Based on the PF results, the dynamic mechanisms of solidification are investigated, including the planar to cellular transition [15,17,31,32], the selection of growth direction [33-35], the competitive growth [36-39], the columnar to equiaxed transition [40,41], and the sidebranching dynamics [42-46], etc. The PF simulations agree well with the experimental observations, indicating its capability of investigating the solidification dynamics. In a word, the PF model avoids the limitations caused by the simplifications in analytical models, which is suitable for investigating the planar instability. Moreover, since it can capture complex morphologies and characteristic parameters of interface, the PF model could reflect the interplay between the interfacial processes and transport processes accurately, which can be used for investigating the planar instability systematically.

In this paper, the dynamics of planar instability in the directional solidification is investigated. Firstly, the interfacial evolution at the initial growth stage is carried out by the WL model and PF model. Secondly, to represent the history-dependence of solidification, constant G and varying $V_P$ are used in the simulations. The solidification evolution under the different conditions are discussed, demonstrating the effect of solute segregation on the excess free energy and interface energy. Finally, to verify the effect of solute segregation, the solidification processes with different Preferred Crystallographic Orientation (PCO) of the crystal are performed. Based on the results, the rules of maximum surface energy and minimum surface stiffness are discussed, indicating the influence of the PCO on the interface instability.

## 2. Models and methodology

### 2.1. Theoretical model

The theoretical model in current paper is based on the linear instability analysis under non-steady-state conditions, the detailed derivations can be found in reference [11]. Here we just present the key equations.

Assuming the pulling direction of the interface is along the z axis. With local equilibrium assumption at the interface, the concentration field ahead of the planar front is:

$$c_0(z_0, t) = -\frac{G}{m} \cdot z_0 \tag{1}$$

where G is the thermal gradient, and $m$ is the slope of liquidus line in the phase diagram. $z_0$ is the position of interface, based on which the instantaneous velocity of interface is defined as [11]:

$$V_I = V_P(t) + \frac{\partial z_0}{\partial t} \tag{2}$$

where $V_P$ is the time-dependent pulling speed. At the interface, solute should satisfy the conservation law:

$$-D_L \frac{\partial c_0}{\partial z}\bigg|_{z0} = V_I(1-k) c_0(z_0, t) \tag{3}$$

where $D_L$ are the solute diffusion coefficient in liquid, and k is the solute partition coefficient. Furthermore, at the planar growth stage, the time-dependent concentration can be approximated by the expression [11]:

$$c_0(z, t) = c_\infty + [c_0(z_0, t) - c_\infty] \cdot \exp\left[\frac{-2(z - z_0)}{l}\right] \tag{4}$$

where $c_\infty$ is the average concentration, and $l$ is the diffusion length. The partial derivative of equation (4) is:

$$\frac{\partial c_0}{\partial z} = \left[c_0(z_0,t) - c_\infty\right] \cdot \exp\left[\frac{-2(z-z_0)}{l}\right] \cdot \left(\frac{-2}{l}\right) \qquad (5)$$

Taking equation (5) to equation (3), we have:

$$\left.\frac{\partial c_0}{\partial z}\right|_{z_0} = \left[c_0(z_0,t) - c_\infty\right] \cdot \left(\frac{-2}{l}\right) = \frac{V_I(1-k)c_0(z_0,t)}{-D_L} \qquad (6)$$

Then, the time-dependent concentration field at the interface can be expressed as:

$$c_0(z_0,t) = \frac{2D_L c_\infty}{2D_L - V_I(1-k)l} \qquad (7)$$

On the other hand, the time-dependent $z_0$ and $l$ can be expressed as [11]:

$$\frac{\partial z_0}{\partial t} = V_I - V_P(t) = \frac{2D_L(z_0 - z_\infty)}{l(1-k)z_0} - V_P(t) \qquad (8)$$

$$\frac{\partial l}{\partial t} = \frac{4D_L(z_\infty - kz_0)}{l(1-k)z_0} - \frac{l}{z_0 - z_\infty}\frac{\partial z_0}{\partial t} \qquad (9)$$

where $z_\infty$ is the steady-state position of the planar interface with the relation of $z_\infty = -m \cdot c_\infty / G$.

By solving equations (7-9), we can obtain the position of interface under non-steady-state conditions, as well as the characteristic parameters, including the solute concentration, instantaneous interface velocity and diffusion length ahead of the planar front.

For a small time interval $\Delta t$, it is easy to show the relationships in equations (10-11) [11], which can be regarded as the initial conditions when solving equations (7-9) numerically.

$$l \approx \left(\frac{8D_L \cdot \Delta t}{3}\right)^{1/2} \qquad (10)$$

$$z_0 = z_\infty - V_P(t) \cdot \Delta t + \frac{V_P(t)\sqrt{2D_L}}{\sqrt{3} \cdot z_\infty(1-k)}(\Delta t)^{3/2} \qquad (11)$$

## 2.2. Phase-Field model

### 2.2.1. Description of the model

The detailed derivation and validation of the quantitative PF model for alloy solidification can be found in literatures [25-27,47]. Here we just present the equations describing the phase field and solute field.

Firstly, a scalar variable $\phi(\mathbf{r}, t)$ is introduced to identify the phase, where $\phi = +1$ reflects the solid phase, $\phi = -1$ reflects the liquid phase, and intermediate values of $\phi$ correspond to the S/L interface. Since $\phi$ varies smoothly across the interface, the usual sharp interface becomes diffuse, and the phases turn into a continuous field, i.e., the phase field.

For the solute field, the composition c(**r**, t) is represented via the supersaturation field U(**r**, t):

$$U = \frac{1}{1-k}\left(\frac{2kc/c_\infty}{1+k-(1-k)\cdot\phi}-1\right) \tag{12}$$

For the alloy solidification, the introduction of the ATC could recover the local equilibrium at the S/L interface and eliminate the spurious effects when the interface width is larger than the capillary length [25]. The ATC term with solute diffusion in solid is given by [47,48]:

$$\vec{j}_{at} = -\frac{1-k\cdot D_S/D_L}{2\sqrt{2}}\left[1+(1-k)U\right]\frac{\partial\phi}{\partial t}\frac{\vec{\nabla}\phi}{|\vec{\nabla}\phi|} \tag{13}$$

where $D_S$ and $D_L$ are the diffusion coefficients in the solid and liquid, respectively. $\partial\phi/\partial t$ reflects the rate of solidification, $\nabla\phi/|\nabla\phi|$ is the unit length along the normal direction of S/L interface.

For the cubic crystal Al-Cu alloy, a four-fold anisotropy function in 2D system is used in this paper:

$$a_s(\hat{n}) \equiv a_s(\theta+\theta_0) = 1+\varepsilon_4\cos 4(\theta+\theta_0) \tag{14}$$

where $\varepsilon_4$ is the anisotropy strength, $\theta$ the angle between the normal direction of S/L interface and the z-axis, $\theta^0$ is the intersection angle between the PCO of grain and the z-axis.

For directional solidification, the "frozen temperature approximation" is adopted, shown in equation (15). The approximation is on the basis of the assumptions: (1) The latent heat is ignored, i.e., the temperature field is undisturbed by the evolution of the S/L interface. (2) There is no flow in the liquid, consistent with the assumption that the densities of the solid and liquid are equal [5].

$$T(z,t) = T_0 + G(t)\left(z-z_0-\int V_P(t)dt\right) \tag{15}$$

Finally, the governing equations of phase field and supersaturation field are given by [25-27]:

$$a_s^2(\hat{n})\left[1-(1-k)\frac{z-z_0-\int V_P(t)dt}{l_T}\right]\frac{\partial\phi}{\partial t} = \\ \nabla\cdot\left[a_s^2(\hat{n})\vec{\nabla}\phi\right]-\partial_x\left(a_s(\hat{n})\cdot a_s'(\hat{n})\cdot\partial_y\phi\right)+\partial_y\left(a_s(\hat{n})\cdot a_s'(\hat{n})\cdot\partial_x\phi\right) \\ +\phi(1-\phi^2)-\lambda(1-\phi^2)^2\left[U+\frac{z-z_0-\int V_P(t)dt}{l_T}\right] \tag{16}$$

$$\left(\frac{1+k}{2}-\frac{1-k}{2}\phi\right)\frac{\partial U}{\partial t}=\nabla\cdot\left[\overline{D}_L\cdot q(\phi)\cdot\vec{\nabla}U-\vec{j}_{at}\right]+\frac{1}{2}\left[1+(1-k)U\right]\frac{\partial\phi}{\partial t} \qquad (17)$$

where,

$$l_T=\frac{\Delta T_0}{G(t)}=\frac{|m|c_\infty(1-k)}{kG(t)}$$

$$\overline{D}_L=D_L/(W^2/\tau_0)$$

$$q(\phi)=\left[kD_S+D_L+k(D_S-D_L)\phi\right]/2D_L$$

In the equations, $l_T$ is the thermal length, where $m$ is the slope of liquidus line in the phase diagram. $q(\phi)$ is an interpolation function determining the varied diffusion coefficient across the domain. Neglecting the influence of kinetic undercooling, the calculation parameters in the PF equations can be linked to physical qualities by expressions $W = d_0\lambda/a_1$ and $\tau_0 = a_2\lambda W^2/D_L$, where $W$ and $\tau_0$ reflect the interface width and relaxation time, which are the length scale and time scale, respectively. In the expressions, $a_1 = 5\sqrt{2}/8$, $a_2 = 47/75$, $\lambda$ is the coupling constant, $d_0 = \Gamma/|m|(1-k)(c_\infty/k)$ is the chemical capillary length. $\Gamma = \gamma_{sl}T_f/(\rho_s L_f)$ is the Gibbs-Thomson coefficient, where $\gamma_{sl}$ is surface energy between solid and liquid, $T_f$ is the melting point of pure solvent and $L_f$ is the latent heat, respectively.

### 2.2.2. Simulation and parameters

The material parameters of Al-2.0wt.%Cu, regarded as a dilute binary alloy, are shown in Table 1 [49,50].

Table 1. The material parameters of Al-2.0wt.%Cu for the simulation [49,50]

| Symbol | Value | Unit |
|---|---|---|
| Liquidus temperature, $T_L$ | 927.8 | K |
| Solidus temperature, $T_S$ | 896.8 | K |
| Diffusion coefficient in liquid phase, $D_L$ | $3.0\times10^{-9}$ | m$^2$/s |
| Diffusion coefficient in solid phase, $D_S$ | $3.0\times10^{-13}$ | m$^2$/s |
| Equilibrium partition coefficient, k | 0.14 | / |
| Alloy composition, $c_\infty$ | 2.0 | wt.% |
| Liquidus slope, m | -2.6 | K/wt.% |
| Gibbs-Thomson coefficient, $\Gamma$ | $2.4\times10^{-7}$ | K·m |
| Anisotropic strength of surface energy, $\varepsilon_4$ | 0.01 | / |

In the PF simulation, the most important calculation parameter is the interface width W [51]. The accuracy of simulation increases with the decrease of W, while the computational cost increases dramatically with the decrease of W. By implementing the thin interface limitation, the value of W in this PF model just needs to be one order of magnitude smaller than the characteristic length scale of microstructures [26,52]. Since the characteristic length of alloy solidification is $L_C \sim \sqrt{d_0 * D_L / v_{tip}}$ [5], W was set to be 0.16μm in this paper. During the computation, the periodic boundary conditions were loaded for the phase field and supersaturation field along the Thermal Gradient Direction (TGD). The time step size was chosen below the threshold of numerical instability for diffusion equation, i.e., $\Delta t < (\Delta x)^2/(4D_L)$. This study used fixed grid size $\Delta x = 0.8W$ and time step size $\Delta t = 0.012\tau_0$.

Moreover, to consider the infinitesimal perturbation of thermal noise on the S/L interface, a fluctuating current $J_U$ is introduced to the diffusion equation. By using the Euler explicit time scheme, we have:

$$U^{t+\Delta t} = U^t + \Delta t \left( \partial_t U - \vec{\nabla} \cdot \vec{J}_U \right) \tag{18}$$

During the numerical simulation, the discretized noise in 2D becomes [43,53]:

$$\vec{\nabla} \cdot \vec{J}_U \approx \left( J^n_{x,i+1,j} - J^n_{x,i,j} + J^n_{y,i,j+1} - J^n_{y,i,j} \right) / \Delta x \tag{19}$$

The components of $J_U$ are random variables obeying a Gaussian distribution, which has the maximum entropy relative to other probability distributions [42]:

$$\left\langle J^m_U(\vec{r},\vec{t}) J^n_U(\vec{r}\,',\vec{t}\,') \right\rangle = 2D_L q(\psi) F^0_U \delta_{mn} \delta(\vec{r}-\vec{r}\,') \delta(t-t') \tag{20}$$

In addition, the constant noise magnitude $F_u^0$ is defined as [43,53]:

$$F^0_U = \frac{k v_0}{(1-k)^2 N_A c_\infty} \tag{21}$$

$F_U^0$ is the value of $F_U$ for a reference planar interface at temperature $T_0$, where $v_0$ is molar volume of the solute atoms, and $N_A$ is the *Avogadro* constant.

Finally, the program code of PF simulation was written by C++ and executed on the platform of π 2.0 cluster, supported by the Center for High Performance Computing at the Shanghai Jiao Tong University (SJTU). The explicit Finite Difference Method (FDM) was used for solving the governing equations, and the Message Passing Interface (MPI) parallelization was used for improving the computational efficiency.

## 3. Results and discussion

In directional solidification, the temperature decreases at the front of the interface, solidification takes place to maintain the local thermodynamic equilibrium. At the initial stage, the interface keeps planar and advances slowly to the liquid. As time goes on, the planar instability appears, which influences the subsequent solidification stages significantly, which deserves systematic investigation.

The dynamic evolution of the interface at the planar instability stage, as well as the influences of solute segregation and the PCO of crystal on the planar instability, will be discussed in the following.

### 3.1. Dynamic evolution at the planar growth stage

In this section, the theoretical WL model and PF model are used to simulate the crystal growth, for which the thermal gradient G is $10^5$K/m and pulling speed $V_P$ is 300μm/s. To represent the same conditions in the WL model and PF model, the PCO in the PF simulation is set parallel with the TGD, i.e., $\theta_0 = 0°$. The computational domain of the PF simulation is 2000×2000 grids, corresponding to 256.0μm×256.0μm in the real unit. It takes about 20 hours using 40 cores to finish one job.

The following information of the characteristic parameters is from the results from the WL model and PF model, including the solute concentration at the front of interface, the instantaneous velocity of interface, and the diffusion length ahead of planar front. As shown in Figure 1(a), the WL model and PF model show good agreement with each other, validating the accuracy of the PF model. However, after the crossover time of planar instability, the results from the two models differ from each other, due to the complex morphologies of interface after the instability. Since the solutions of the theoretical model could only apply under the simple conditions, i.e., 1D system, the analyses before the planar instability are based on both the WL model and PF model, while the analyses after the planar instability are just based on the PF simulations.

As shown in Figure 1(a), the instantaneous velocity of interface increases with time, so does the solute concentration ahead of the interface. The corresponding interfacial evolution are shown in Figure 2. As time goes on, the planar instability appears, represented by the transition from the planar to the cellular, as shown in Figure 2(c)-(d). Due to the interface shape, the cellular appearing increases the tip velocity, shown by the sharp increment of the orange curve near the crossover time in Figure 1(a). The tip velocity here is defined as $v_{tip} = [z(t_2)-z(t_1)]/(t_2-t_1)$. On the other hand, the solute at the S/L interface should satisfy the conservation law. At the crossover time of planar instability, the solute still accumulates ahead of the interface, shown by the limited increment of the red curve near the crossover time in Figure 1(a). After the cellular appearing, rather than diffusing only along the pulling direction of the planar interface, the solute can diffuse along multiple directions from the cellular tip to the liquid. As a result, the solute concentration starts to decrease, shown by the decrease of the red curve after the peak in Figure 1(a).

To represent the solutal evolution more clearly, we obtain the solute concentration across the interface at the initial planar stage, in Figure 1(b), where the time intervals between each two curves are the same. The peaks of the curve in Figure 1(b) represent the positions of S/L interface, resulting from the solute trapping at interface during alloy solidification. The tendency of the peaks illustrates the solute ahead of the interface firstly increases and then decreases, consistent with the previous discussion. Since the curve peaks represent the positions of interface, the space intervals between the peaks, i.e., the distance between them along the x-axis, represent the pulling distance of the interface at a given time, which reflects the tip velocity. As shown in Figure 1(b), after the crossover time of planar instability (t = 0.399s), the increasing distance between the peaks illustrates the increasing tip velocity, resulting from the transformation from the planar to the cellular.

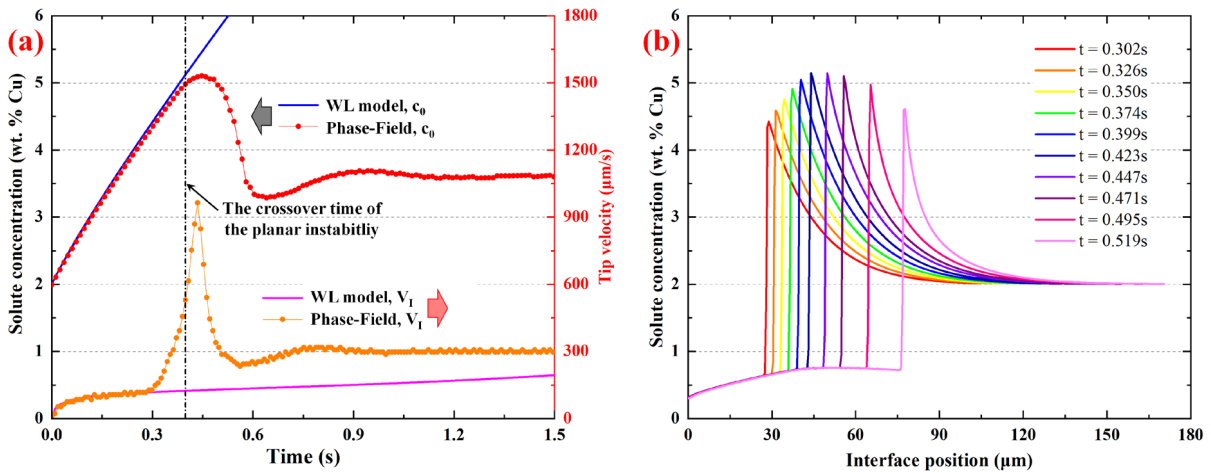

Figure 1. The evolution of the characteristic parameters: (a) the evolution of the solute concentration ahead of the interface and the instantaneous velocity of interface (from the WL model and PF model); (b) the solute concentration across the interface at the initial instability stage (from the PF simulations)

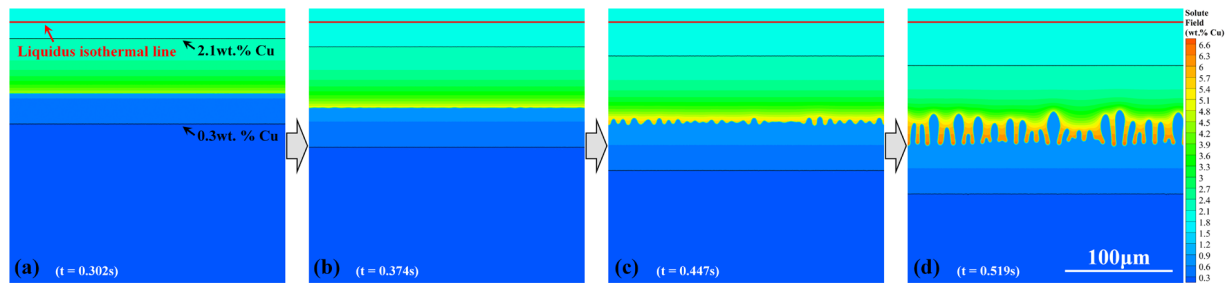

Figure 2. The evolution of interfacial morphology and solute field at the planar instability stage (from the PF simulations)

In conclusion, the results indicate the phenomenon consistently: at the initial stage, both the tip velocity and solute concentration ahead of the interface increase with time. As time goes on, the planar instability occurs, represented by the transformation from the planar to the cellular.

**3.2. The role of solute concentration in the interface instability**

To investigate the history-dependence of solidification, we set G constant while $V_P$ varying with time. The increase times of $V_P$ are 0.0s, 0.5s and 1.0s, respectively, shown in Figure 3(a1)-(a3). The corresponding evolution is shown in Figure 3(b1)-(b3), including the solute concentration and tip velocity. With different increase rates of $V_P$, the evolution of the characteristic parameters show similar tendencies with each other, whereas the quantitative evolution differ with each other. Under the constant G and $V_P$, in Figure 3(b1), the acceleration rate of tip velocity decreases with time, so does the increase rate of solute concentration. That is, the tip velocity and solute concentration increase logarithmically with time, consistent with literature [12]. By contrast, under the constant G and varying $V_P$, in Figure 3(b2)-(b3), the acceleration rate of tip velocity increases gradually at the initial transient stage, so does the increase rate of solute concentration. That is, the tip velocity and solute concentration increase exponentially with time, consistent with literature [17,53].

The results demonstrate the distinctions between the steady-state and dynamic conditions, since it takes time to adjust the parameters in practice, it is more suitable to adopt the dynamic conditions.

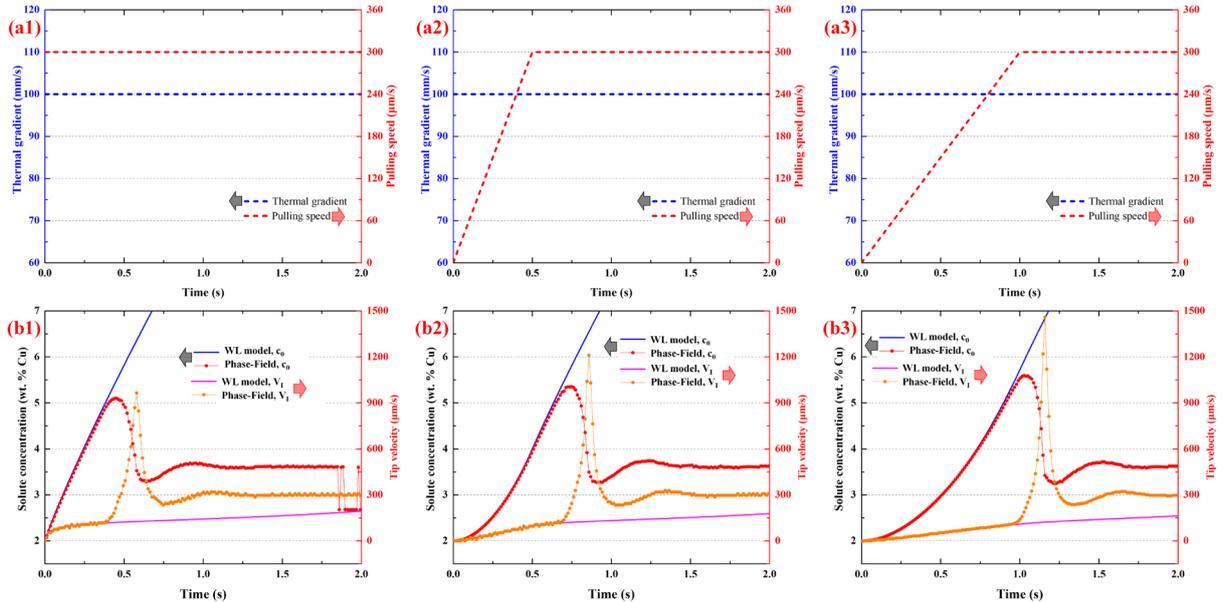

Figure 3. (a) The constant G and varying $V_P$ with different increase rates (the increase times of $V_P$ are 0.0s, 0.5s and 1.0s, respectively); (b) The corresponding evolution of the characteristic parameters, including the solute concentration ahead of the interface $c_0$ and the instantaneous velocity of interface $V_I$.

In the following discussion, we use the dynamic solidification parameters in the simulation. The thermal gradient G is constant while the pulling speed $V_P$ increases with time first and then reaches a fixed value, for which the increase times of $V_P$ are 0.5s, 1.0s, 2.0s and 3.0s, respectively, shown in Figure 4(b).

The corresponding evolution are shown in Figure 4(a), including the solute concentration at the front of interface and tip velocity. Similar with the previous, at the initial stage, both the tip velocity and solute concentration increase with time. As time goes on, the planar instability occurs, represented by the transition from the planar to the cellular, shown by the sharp increments of the orange curves ($v_{tip}$) at the crossover time in Figure 4(a1)-(a4). On the other hand, the solute concentration starts to decrease after the cellular appearing, shown by the red curves after the crossover time.

Despite the same tendencies between the characteristic parameters, under different dynamic conditions, the quantitative evolution differ with each other. As the acceleration rate of $V_P$ increases, from Figure 4(a1) to Figure 4(a4), the incubation time of planar instability increases, shown by the different crossover times. Moreover, the planar instability could occur before $V_P$ becomes constant, in Figure 4(a3), (b3) and (a4), (b4). That is, under the dynamic conditions, the onsets of planar instability correspond to different $V_P$, indicating the G and $V_P$ are not the critical parameters for the interface instability.

Here, rather than the critical thermal parameters G and $V_P$, we attribute the interface instability to the solute concentration, as well as the corresponding interface energy and its anisotropy, as following:

During solidification, the atoms at the S/L interface need to accommodate the slight structural changes on both solid and liquid sides, bringing the excess free energy. The integral of the excess free energy of the interface, multiplied by molar volume, is the S/L interface energy (unit: J/m$^2$), which is given by [5]:

$$\gamma_{sl} = \frac{1}{V^m} \int \Delta G^m(z) dz \tag{22}$$

Equation (22) shows the interface energy is dominated by the excess Gibbs energy. Since the solution in current PF model is regarded as the ideal solution, there is no interaction between the atoms. The material is regarded as the binary alloy made up of solvent Al and solute Cu. Taking the liquid as the reference state (where $G_{Al}^m$ and $G_{Cu}^m$ are zero), the molar free energies of liquid and solid are given by [5]:

$$G_L^m(X_L, T) = RT\left[X_L \ln X_L + (1 - X_L) \ln(1 - X_L)\right] \tag{23}$$

$$\begin{aligned}G_S^m(X_S, T) = &(1 - X_{Cu})\Delta S_f^{Al}(T - T_f^{Al}) + X_{Cu}\Delta S_f^{Cu}(T - T_f^{Cu}) \\ &+ RT\left[X_S \ln X_S + (1 - X_S) \ln(1 - X_S)\right]\end{aligned} \tag{24}$$

where R is the gas constant, T is the temperature, $X_{Cu}$ is the molar composition of Cu. $\Delta S_f^{Al}$ and $\Delta S_f^{Cu}$ are the entropies of fusion of pure Al and Cu, respectively. $T_f^{Al}$ and $T_f^{Cu}$ are the equilibrium fusion points of Al and Cu, respectively.

In this paper, the latent heat is ignored, meaning $\Delta S_f^{Al} = \Delta S_f^{Cu} = 0$. When solidification takes place, the free energy reduces from $G_L$ to $G_S$. The difference between $G_L$ and $G_S$ is the driving force for solidification. Due to the local equilibrium, the compositions $X_S$ and $X_L$ satisfy the relationship $X_S = k \cdot X_L$. According to equations (23-24), the expression of $\Delta G^m$ is:

$$\Delta G^m(X_L, T) = G_L^m(X_L, T) - G_S^m(X_S, T) \\ = RT\left[X_L \ln X_L + (1-X_L)\ln(1-X_L) - k \cdot X_L \ln(k \cdot X_L) + (1-k \cdot X_L)\ln(1-k \cdot X_L)\right] \quad (25)$$

According to equation (25), we know that $\partial \Delta G^m/\partial T < 0$ and $\partial \Delta G^m/\partial X_L < 0$ (when $X_L < 0.5$). That is, $\Delta G^m$ has negative relations with T and $X_L$. That is, the higher temperature and higher degree of the solute segregation correspond to the less excess free energy and lower interface energy. As solidification goes on, T decreases and $\Delta G$ increases, while $X_L$ increases and $\Delta G$ decreases. Since the effect of $X_L$ is much greater than T. The excess free energy $\Delta G^m$ decreases with the solidification time increasing, so does the interface energy. When the interface energy reduces to a critical level, with the influence of interfacial anisotropy, the planar instability appears.

In conclusion, the change of the excess free energy at the interface, caused by the solute segregation, and the corresponding interface energy are the critical parameters of the instability.

Based on the above discussion, during alloy solidification, the solute segregation results from the solute trapping, determined by the propagation speed of interface, i.e., the solidification rate. During the non-rapid solidification, the propagation speed of planar interface is dominated by the cooling rate $R = G*V_P$ (unit: K/s). In Figure 4, the acceleration rate of $V_P$ decreases from (b1) to (b4), corresponding to the decreasing acceleration rate of R, making the solidification rate decrease. As a result, the accumulation rate of solutes decreases, shown by the red curves in Figure 4 from (a1) to (a4). Since the solute segregation determines the interface energy, the decrease rate of the interface energy also decreases, delaying the onset of the planar instability, shown by the different crossover times in Figure 4.

The dynamical evolution corresponding to Figure 4 is shown in Figure 5, including the concentration across the interface, the morphological evolution and solutal evolution. As shown in Figure 5 (a2), (b2), (c2) and (d2), the distance between the instantaneous position of interface and the 0.3wt.%Cu curve corresponds to the pulling distance of interface, which illustrates the pulling distance of the planar interface increases with the decreasing acceleration rate of $V_P$, from (a2) to (d2) in Figure 5. On the one hand, smaller $V_P$ corresponds to smaller cooling rate R, decreasing the propagation speed of interface. On the other hand, the decreasing acceleration rate of $V_P$ delays the onset of planar instability. The pulling distance of the planar interface satisfies the relationship $z_p=\int V_{tip}(t)dt$. Before the crossover time, the integral of the velocity curve versus time reflects the pulling distance of planar interface, shown by the area below the tip velocity curve from t = 0 to the crossover time in Figure 4(a). The increasing areas, from (a1) to (a4) in Figure 4, illustrate the increasing pulling distance of the planar interface, consistent with the result in Figure 5 (a2), (b2), (c2), (d2). Moreover, as shown in Figure 5 (A)-(D), the peak position of the green curves represent the position of interface at the onset of instability. The increasing x-axis positions of the green peaks reflect the increasing pulling distance of planar interface, consistent with the previous discussion.

In conclusion, in alloy solidification, the cooling rate R ( = G*$V_P$) dominates the overall propagation speed of the interface, to maintain the local thermodynamic equilibrium. The solute segregation determines the stability of the interface, by changing the excess free energy and the corresponding interface energy.

It needs to be noted the distance between the position of interface and the solute curve of 2.1wt.%Cu represents the diffusion length ahead of the planar front, as shown in Figure 5(a2), (b2), (c2), and (d2). The diffusion lengths calculated from the WL model and PF model differ from each other, indicating the models need to be modified, which will be carried out in the future.

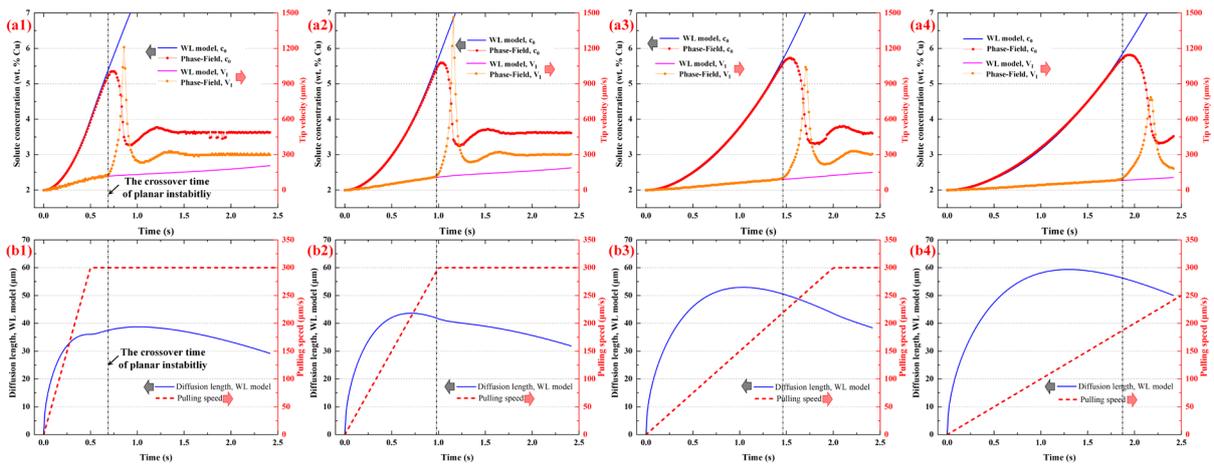

Figure 4. (a) The evolution of the characteristic parameters with different increase rates of $V_P$, including the solute concentration ahead of interface and the instantaneous velocity of interface; (b) the time-dependent $V_P$ and the diffusion length ahead of the planar front.

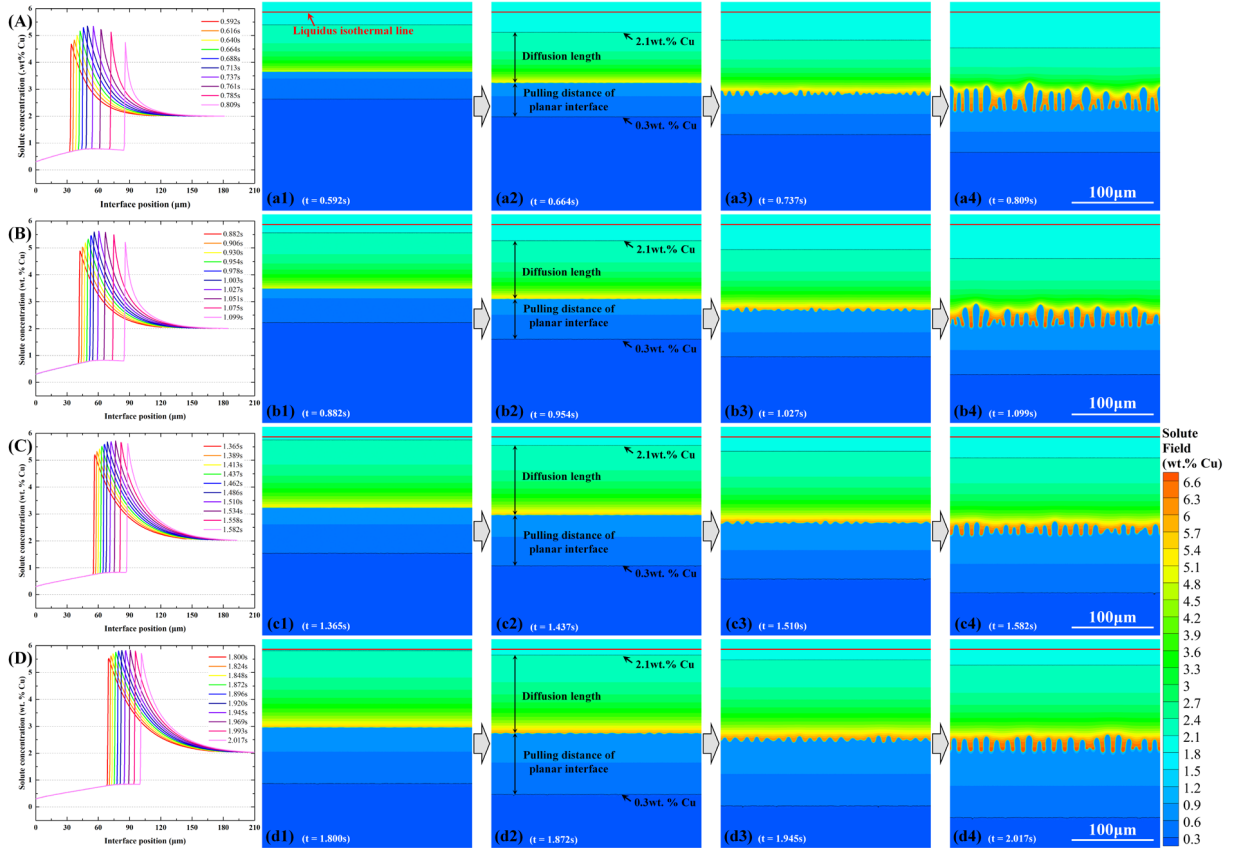

Figure 5. The solute concentration across the interface at different times (A-D), and the evolution of interface morphology and solute field at the planar instability stage (from the PF simulation)

### 3.3. Planar instability with different preferred crystallographic orientations

The previous simulations are based on the setting that the PCO of crystal is parallel with the TGD. In the actual solidification process, the <100> directions of crystal are not always parallel with the TGD, i.e., there is an intersection angle θ₀ between the PCO of the crystal and the TGD. Since the crystals with different PCOs have different interface energy, the simulations with different PCOs can test the conclusion that the solute segregation influencing the excess free energy and the corresponding interface energy.

In a 2D system, the surface energy of the cubic crystal can be expressed by:

$$\gamma_{sl} = \gamma_{sl}^0 \left[1 + \varepsilon_4 \cos 4(\theta + \theta_0)\right] \quad (26)$$

where $\gamma^0_{sl}$ means the value of isotropic surface energy, $\varepsilon_4$ is a measure of the strength of the anisotropy.

The surface stiffness is defined as [54]:

$$\Psi_{sl} = \gamma_{sl} + \frac{d^2 \gamma_{sl}}{d\theta^2} = \gamma_{sl}^0 \left[1 - 15\varepsilon_4 \cos 4(\theta + \theta_0)\right] \quad (27)$$

As shown in equations (26-27), the PCO determines directly the surface energy and surface stiffness. There are two common rules for the selection of growth direction: maximum surface energy and minimum surface stiffness. The initial instability is presented by the transition from the planar to the cellular, where the growth direction of the cellular is dominated by the surface energy and/or surface stiffness [55]. As a result, by adjusting the surface energy and/or surface stiffness, the PCO of crystal will affect the planar instability.

In this section, solidification processes with the different PCOs are carried out, for which the PCOs are set to be 0°, 15°, 30° and 45°, respectively. The computational domain is 2000×2000 grids, corresponding to 256.0μm×256.0μm in the real unit. It takes 20 hours using 40 cores to finish one job.

The evolution of the characteristic parameters are shown in Figure 6, including the solute concentration ahead of the interface $c_0$ and tip velocity $V_I$. The $c_0$ curves overlap with each other completely before the crossover time, so do the $V_I$ curves. The results indicate the surface energy and its anisotropy do not affect the solute diffusion and planar growth process, consistent with the literature [53]. As time goes on, the planar instability appears, shown by the sharp increase of $V_I$ curves in Figure 6, illustrating the crossover times of the instability are almost the same between the simulations with different PCOs. That is, the PCO of crystal has little influence on the crossover time of the planar instability. The results agree well with the discussion in the previous section. The solute segregation decreases the interface energy and promotes the onset of instability. At the planar growth stage, the degree of solute segregation is determined by the cooling rate. Meanwhile the same cooling rates are used in the simulations, resulting in the same degree of the solute segregation. Hence, the decreases of the excess free energy and interface energy are also the same between the simulations, resulting in the same crossover times in Figure 6.

Although the crossover times of planar instability are almost the same between the simulations with different PCOs, the detailed evolution of the planar-cellular-transition (PCT) show differences, represented by the evolution of solute concentration in Figure 6. As mentioned before, the sharp increment of tip velocity reflects the onset time of planar instability, while the peak of solute concentration reflects the completion of the cellular appearing. In Figure 6, the different peak positions of solute concentration curves represent the different times for the PCT, where the amounts of time gradually increases from $\theta_0 = 0°$ to $\theta_0 = 45°$.

The detailed evolution corresponding to Figure 6 are shown in Figure 7. Before the planar instability, in Figure 7(a1), (b1), (c1), and (d1), the degrees of solute segregation are similar between the simulations, corresponding to similar interface energy. As time goes on, the interfacial evolution show the tendency of instability, shown by the fluctuations of the interface in Figure 7(a2), (b2), (c2), and (d2). At this stage, the pulling distance of planar interface and the diffusion length ahead of planar front are also similar. The results indicate, before the planar instability, the PCO of crystal hardly influences the solute diffusion process, whose evolutional characteristics are almost the same between the simulations with different PCOs.

Although the planar growth and the onset times of planar instability are similar between the simulations, the detailed evolution of the PCT show differences, in Figure 7(a3), (b3), (c3), and (d3). The differences are more obvious in Figure 7(a4), (b4), (c4), and (d4), including the spent time of the PCT and the morphology of the cellular. The differences result from the different PCOs of crystal. As mentioned before, there are two common rules for the selection of growth direction: maximum surface energy and minimum surface stiffness. For the cubic crystal, the rule of maximum surface energy means the crystal will seek to minimize the total surface energy by creating higher curvature in the <100> direction, while the rule of minimum surface stiffness means the crystal will prefer to grow in the direction where the surface presents the smallest resistance to being deformed [5]. As for the PCT, it is difficult for the cellular to create high curvature in the <100> direction directly from the planar with zero-curvature, due to the great energy barrier caused by curvature difference. That is, the rule of maximum surface energy is not suitable for the PCT. Then we move on to the rule of minimum surface stiffness, in equation (27), the surface stiffness is expressed by $\Psi_{sl} = \gamma^0_{sl}[1-15\varepsilon_4\cos4(\theta+\theta_0)]$, where $\theta$ is the angle between the normal vector of interface and the z-axis. At the planar growth stage, the value of $\theta$ is infinitesimal. The expression of surface stiffness can be simplified as $\Psi_{sl} = \gamma^0_{sl}[1-15\varepsilon_4\cos4\theta_0]$, based on which we could know the surface stiffness increases from $\theta_0 = 0°$, 15°, 30° to $\theta_0 = 45°$. According to the minimum surface stiffness rule, the crystal will prefer to grow in the direction where the surface presents the smallest resistance to being deformed. Hence, the crystal with $\theta_0 = 0°$ is the easiest to appear, while the crystal with $\theta_0 = 45°$ is the hardest to appear, shown by the different amplitudes of perturbation of interface in Figure 7(a3), (b3), (c3), and (d3). Meanwhile, the growth directions and interface morphologies show few differences between the simulations at the initial stage of the cellular growth, since the curvature of interface increases gradually from zero to the high level, which takes some time.

After the cellular appearing, the stochastic factors make the cellular have different shapes and growth speeds, in Figure 7(a4), (b4), (c4), and (d4), resulting in temporary competition between the cellular. The <100> direction of crystal dominates the growth direction and the interfacial morphology, affecting the solute diffusion and corresponding solute concentration. In turn, the solute concentration influences the excess free energy and interface energy, determining the cellular growth. The evolution of subsequent solidification stages will be discussed in the future.

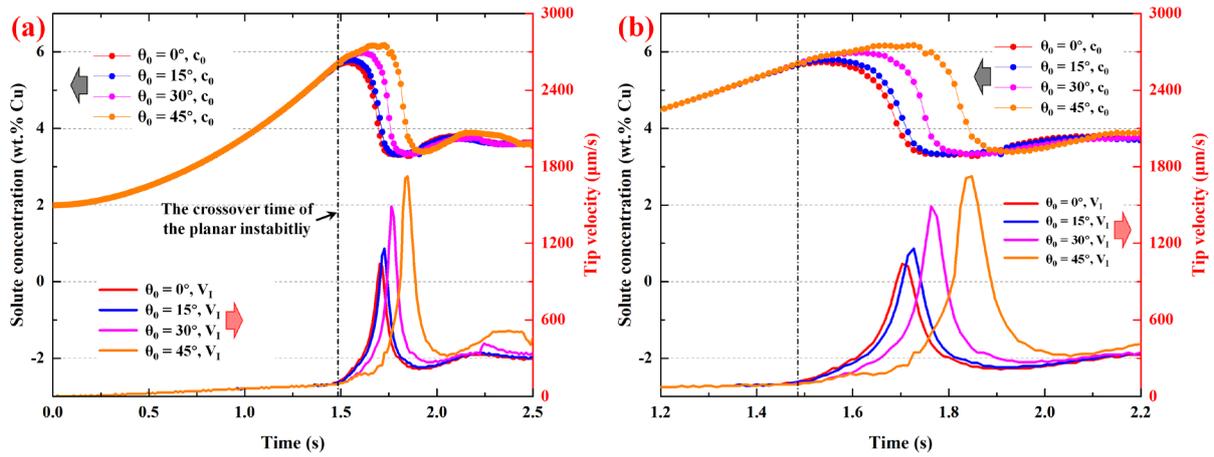

Figure 6. The evolution of the characteristic parameters with time: (a) the solute concentration ahead of the interface and the instantaneous velocity of interface; the enlarged version of (a). (from the PF simulation).

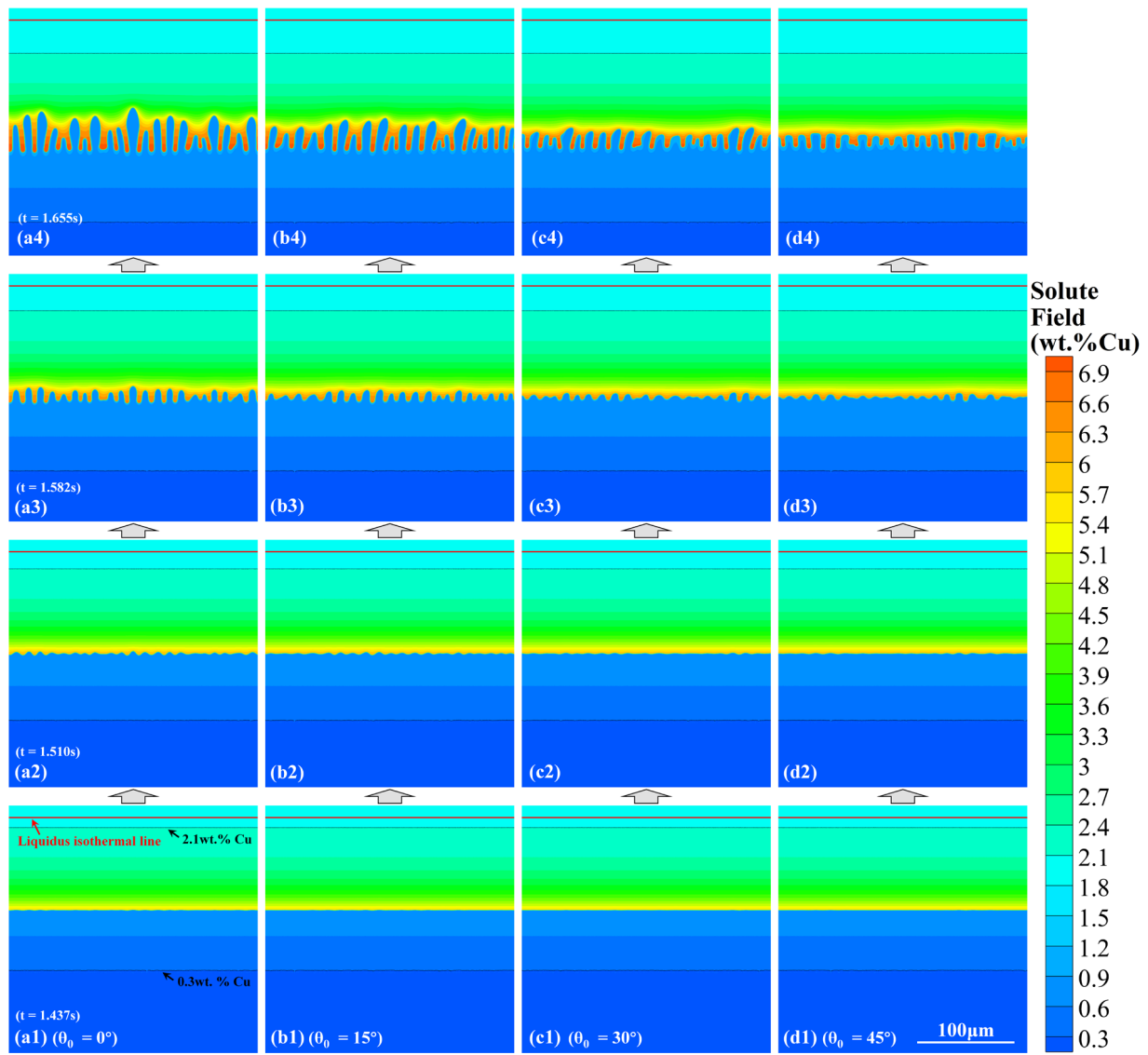

Figure 7. The evolution of interface morphology and solute field with different PCOs at the planar instability stage (from the PF simulation)

In conclusion, the simulations with different PCOs indicate the surface anisotropy does not affect solute diffusion and planar growth. The results also verifies the effect of solute segregation on the interface energy, resulting in the instability. On the other hand, the minimum surface stiffness rule is more suitable than the maximum surface energy rule for the PCT. The curvature increases gradually from zero to the high level at the <100> direction, taking different times between the simulations, due to the different surface stiffness.

## 4. Summary and outlook

We investigate the planar instability in directional solidification. Firstly, the interfacial evolution at the initial growth stage is performed by the WL model and PF model, respectively. Secondly, to represent the history-dependence of solidification, the constant G and varying $V_P$ are used in the simulations. The evolution under different dynamic conditions is discussed, demonstrating the solidification evolution depends on the detailed way the conditions are achieved. In addition, the results indicate the influence of the solute on the interface instability. Finally, the solidification processes with different PCOs of crystal are carried out, the rules of maximum surface energy and minimum surface stiffness are discussed. The following conclusions can be drawn from the study:

(1) At the initial stage, both the tip velocity and solute concentration ahead of the S/L interface increase with time. As time goes on, the planar instability occurs, represented by the transformation from the planar to the cellular.

(2) In alloy solidification, the cooling rate R ( = G*$V_P$) dominates the overall propagation speed of the interface, to maintain the local thermodynamic equilibrium. The solute segregation determines the stability of the interface, by changing the excess free energy and the corresponding interface energy.

(3) The simulations of the crystals with different PCOs indicate the surface energy and its anisotropy do not affect the solute diffusion and planar growth. The results verify the conclusion that the solute segregation influences the interface energy and results in the interface instability. On the other hand, for the planar-cellular transition, the minimum surface stiffness rule is more suitable than the maximum surface energy rule.

The influence of the solute concentration on the excess free energy and interface energy can be applied to other solidification patterns induced by the interface instability, including the morphological selection and the onset of sidebranching, etc, which will be studied in the future.

**Acknowledgments**

This work is supported by the fellowship of China Postdoctoral Science Foundation (2021M692040), National Key Research and Development Program (2017YFB0305301), and National Natural Science Foundation of China-Excellent Young Scholars (51922068). The authors acknowledge the technical support from the Center for High Performance Computing at the SJTU.

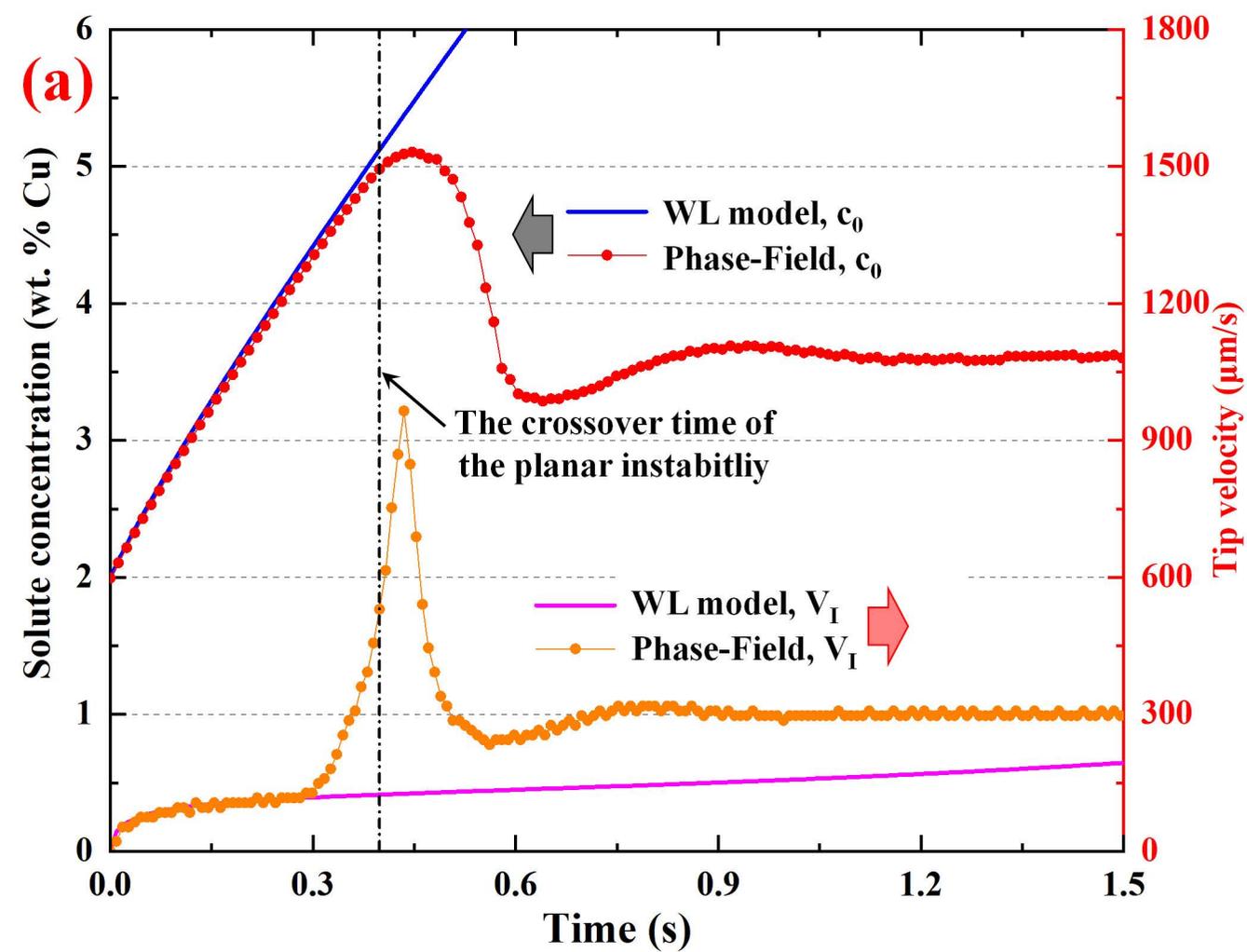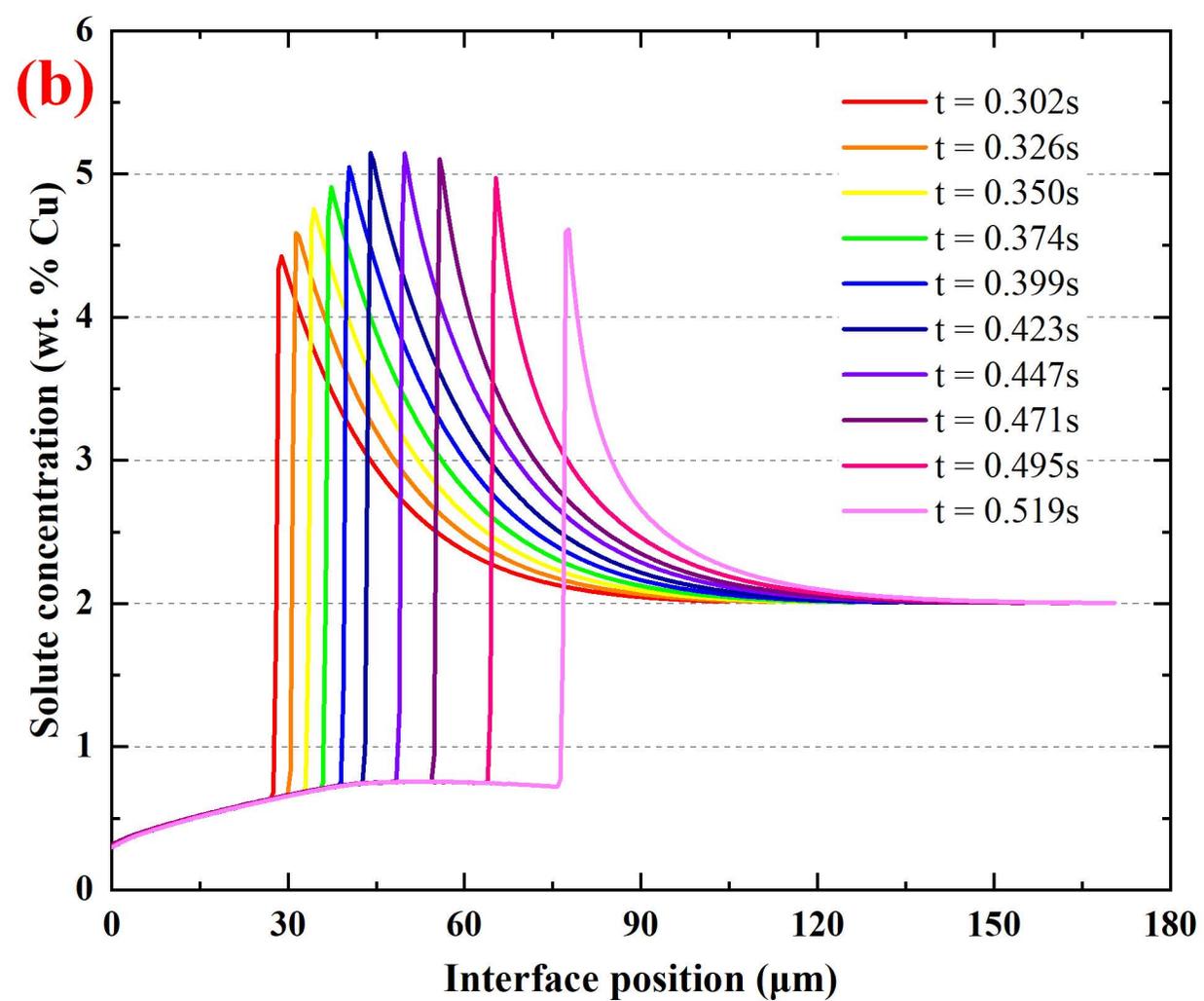

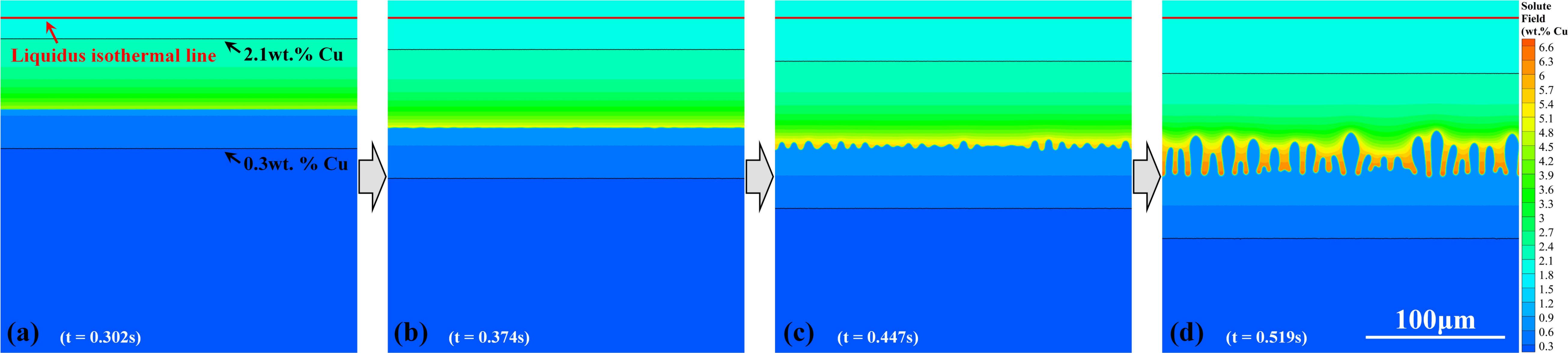

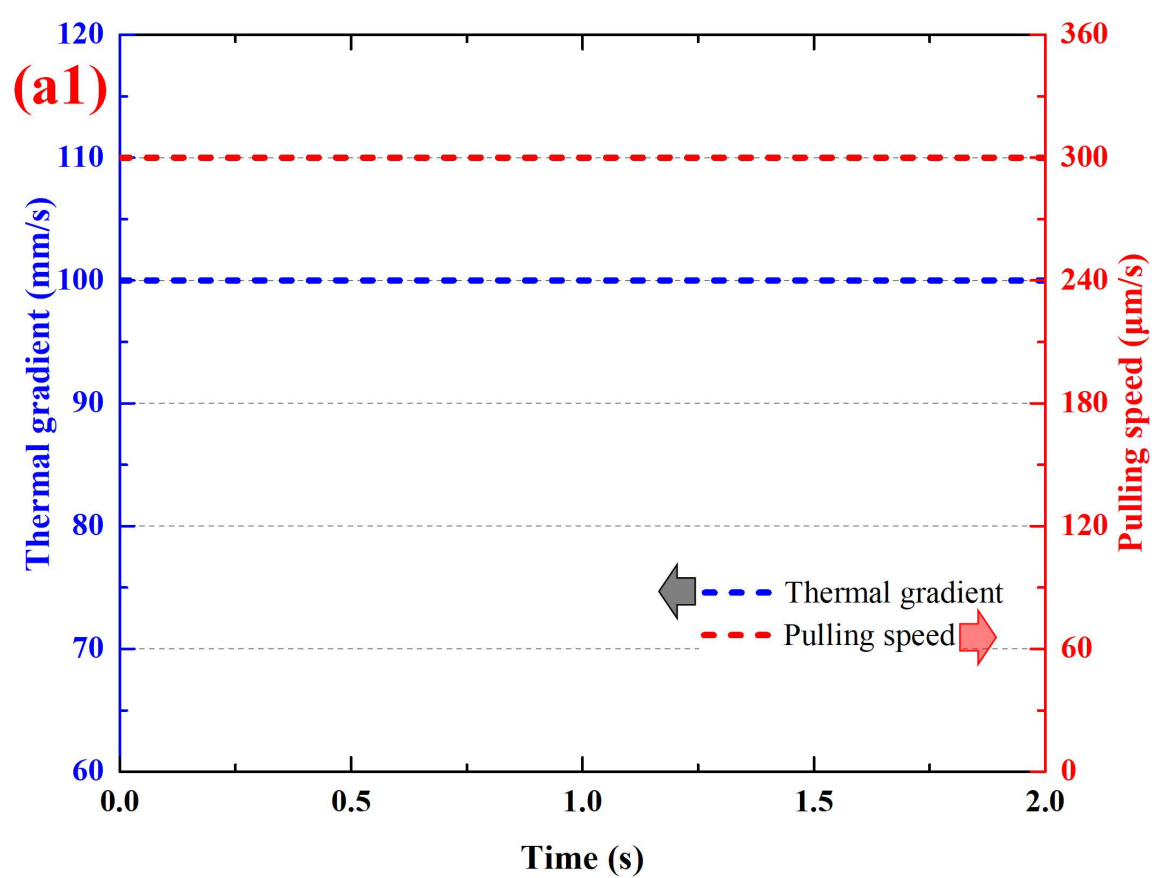
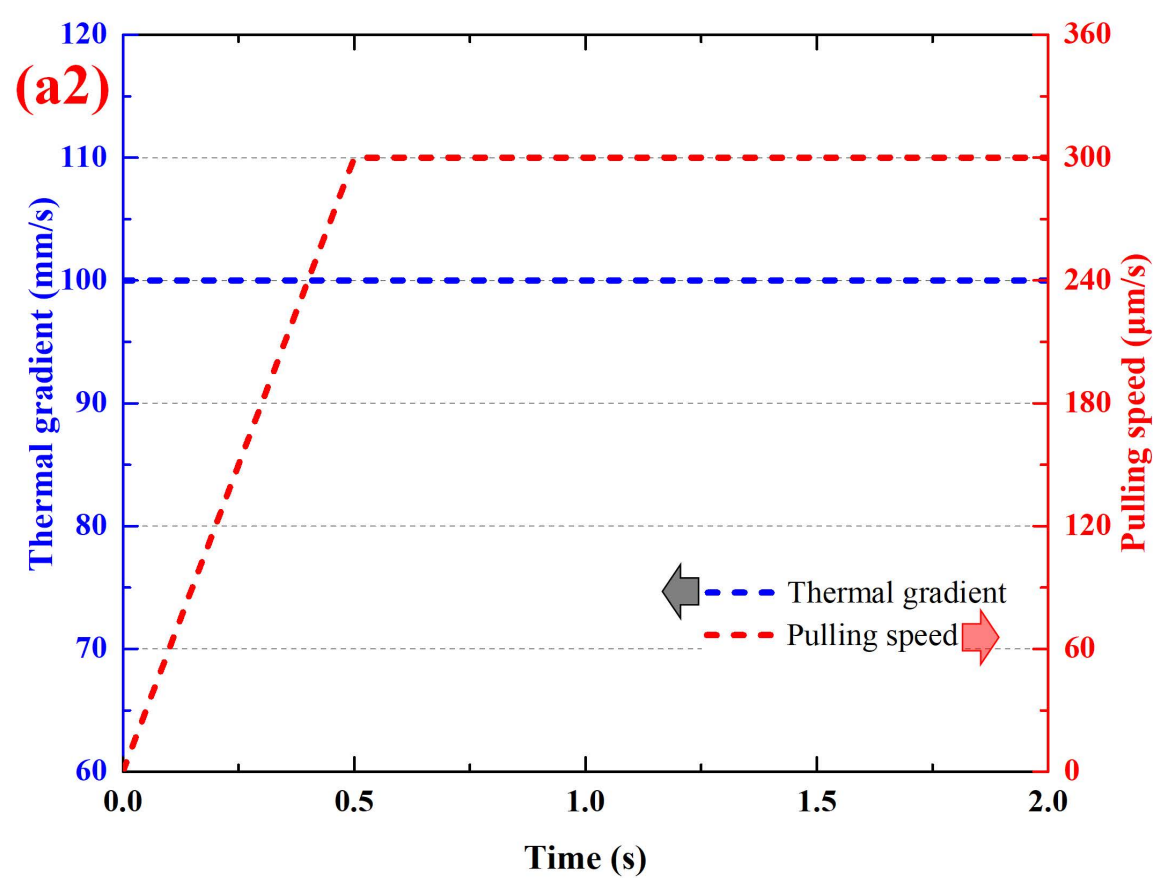
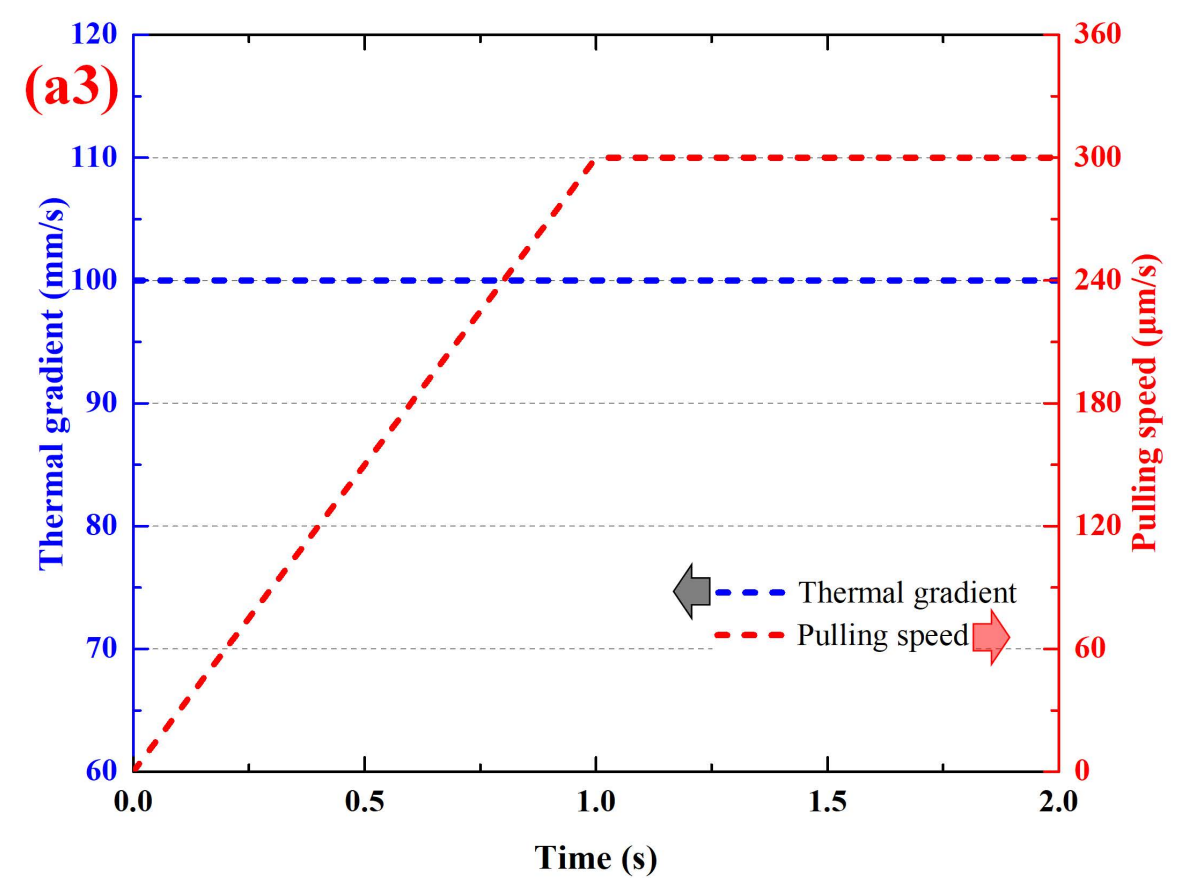
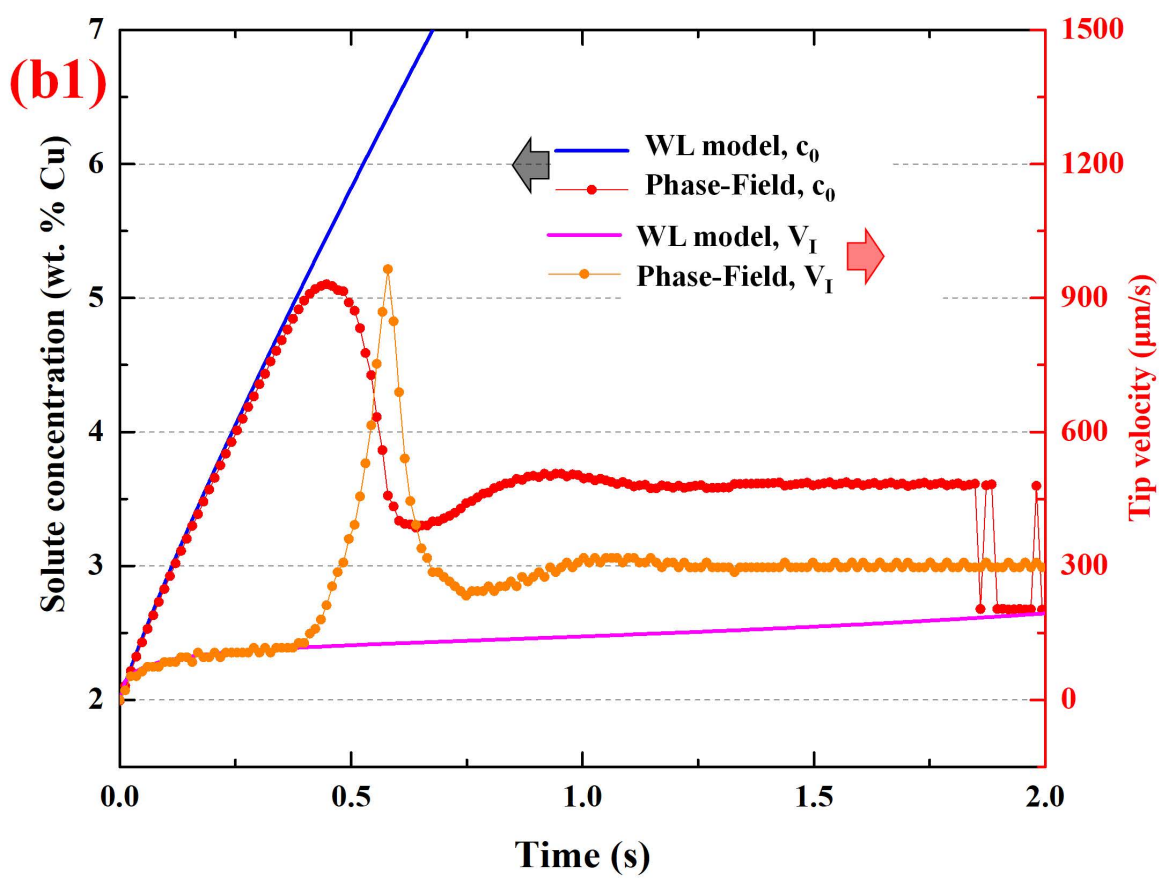
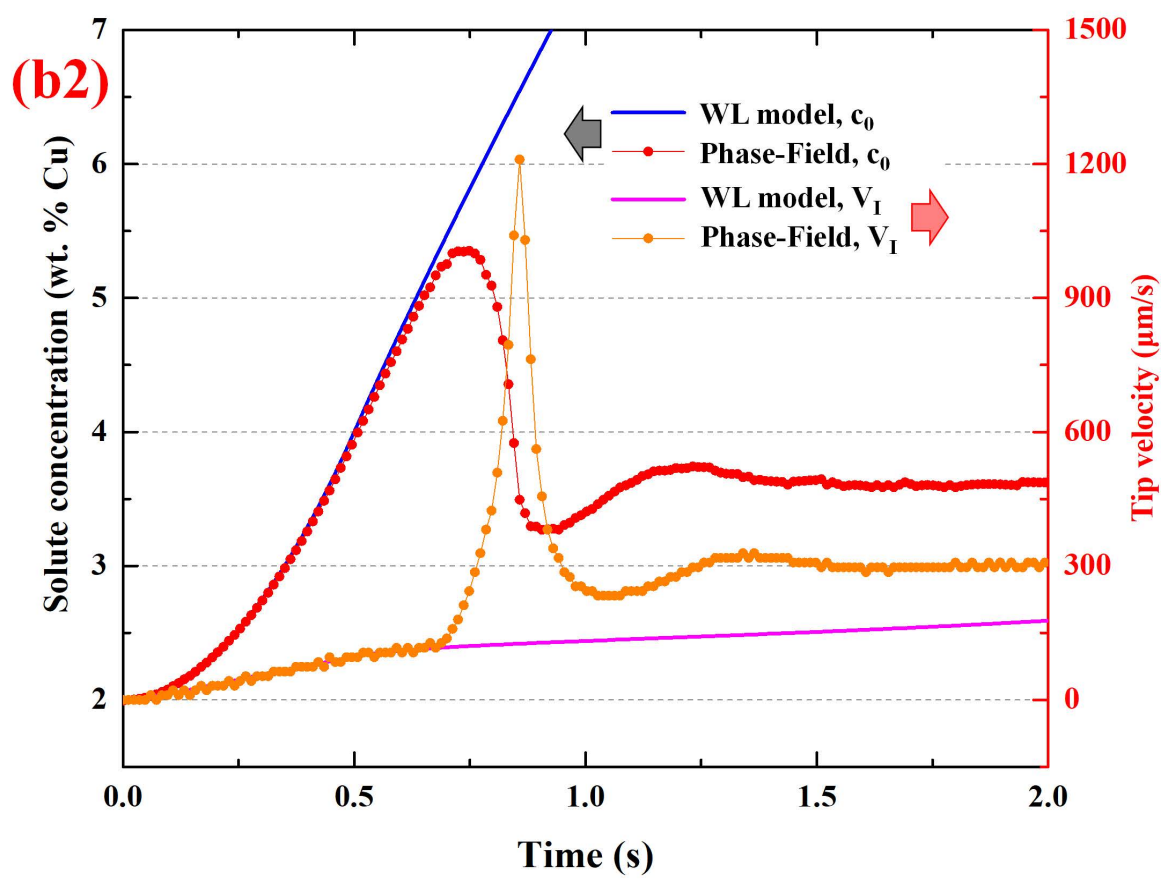
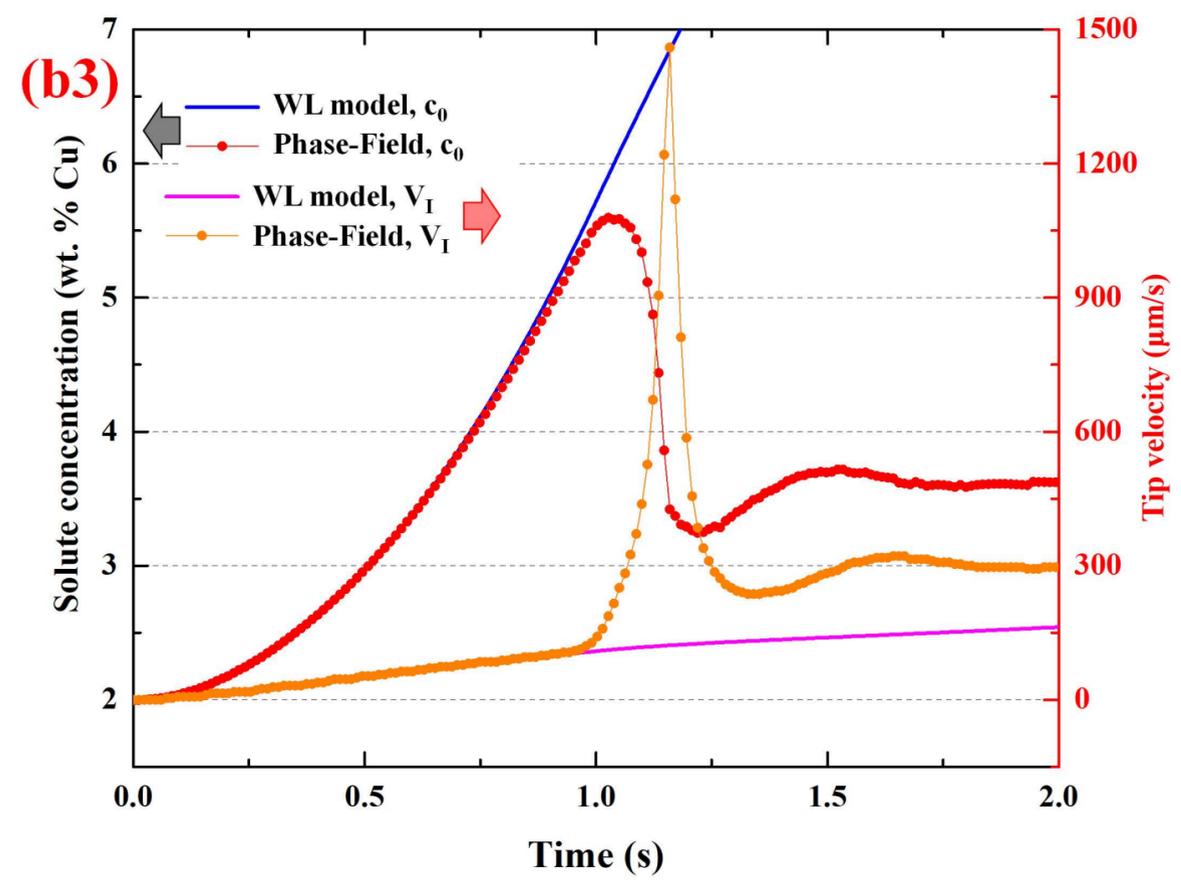

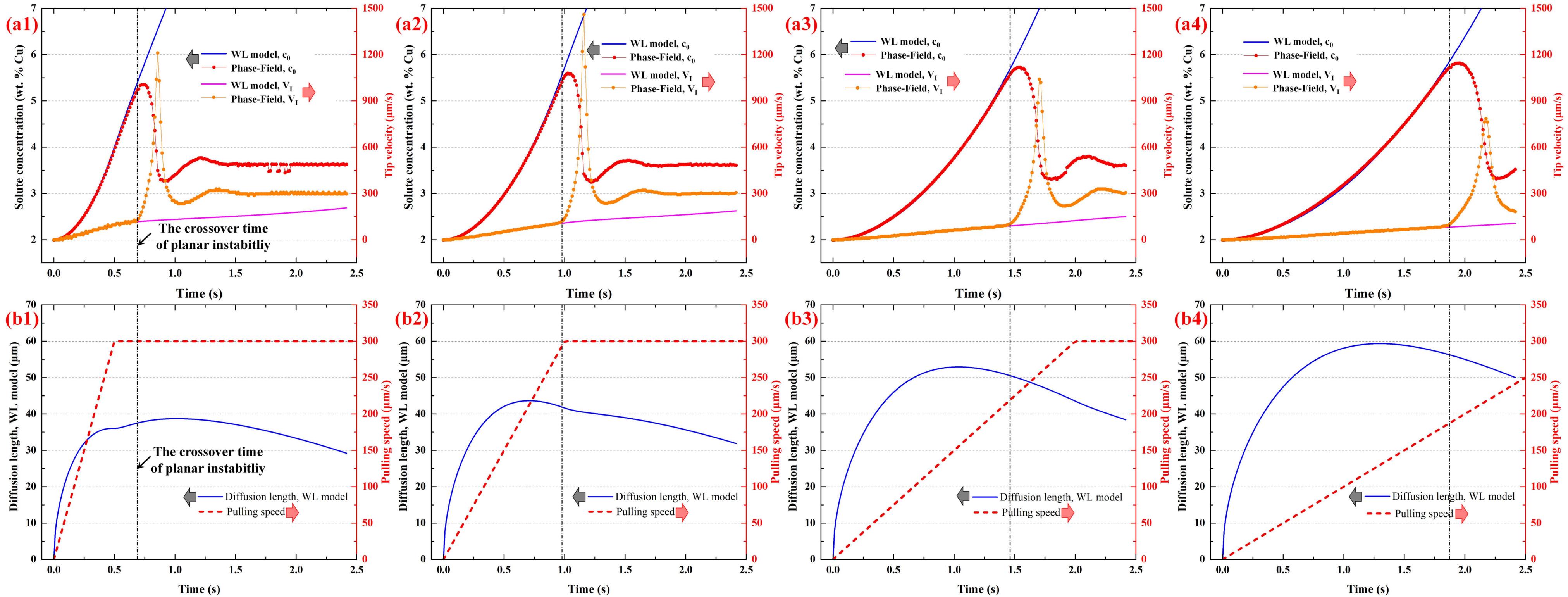

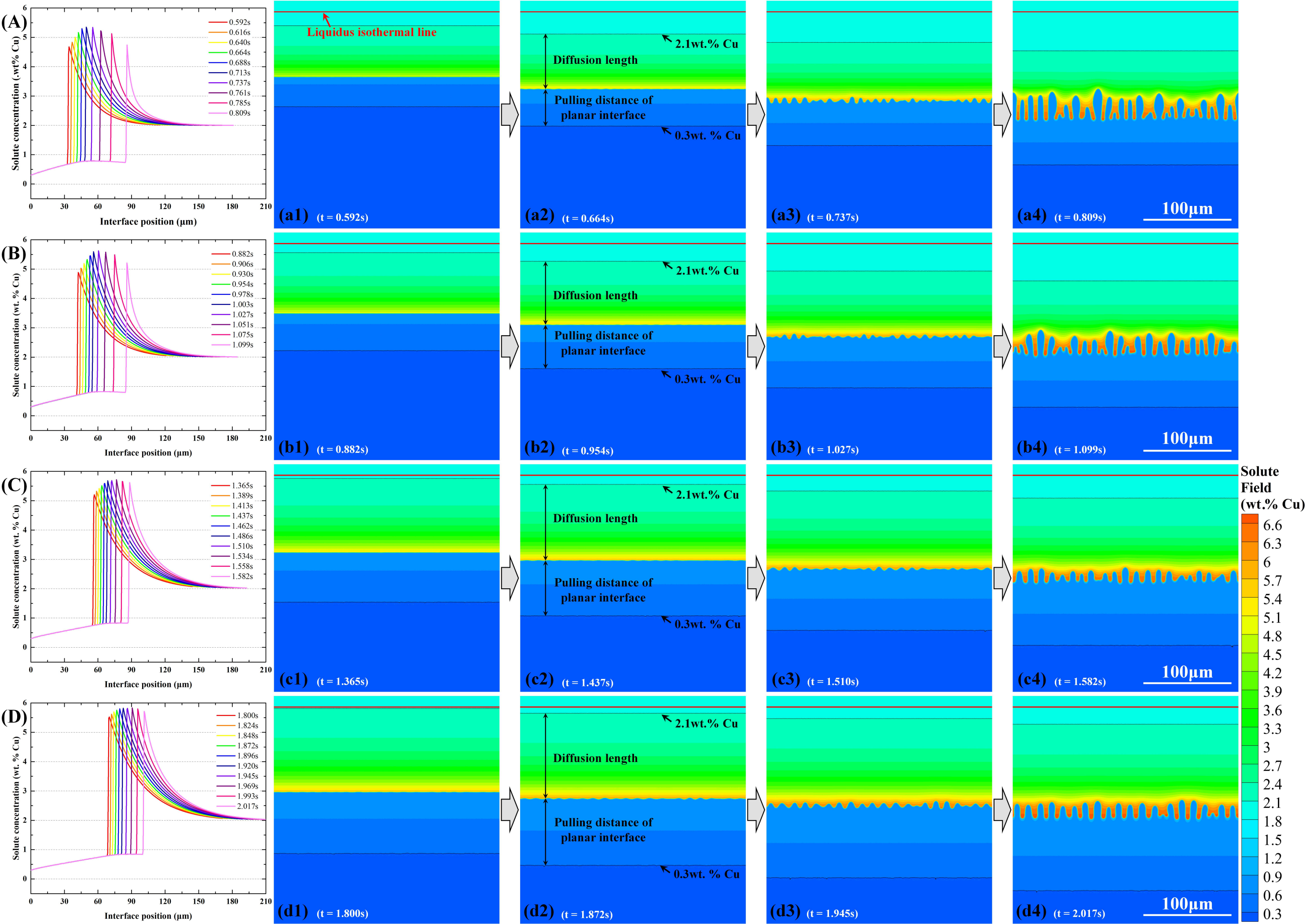

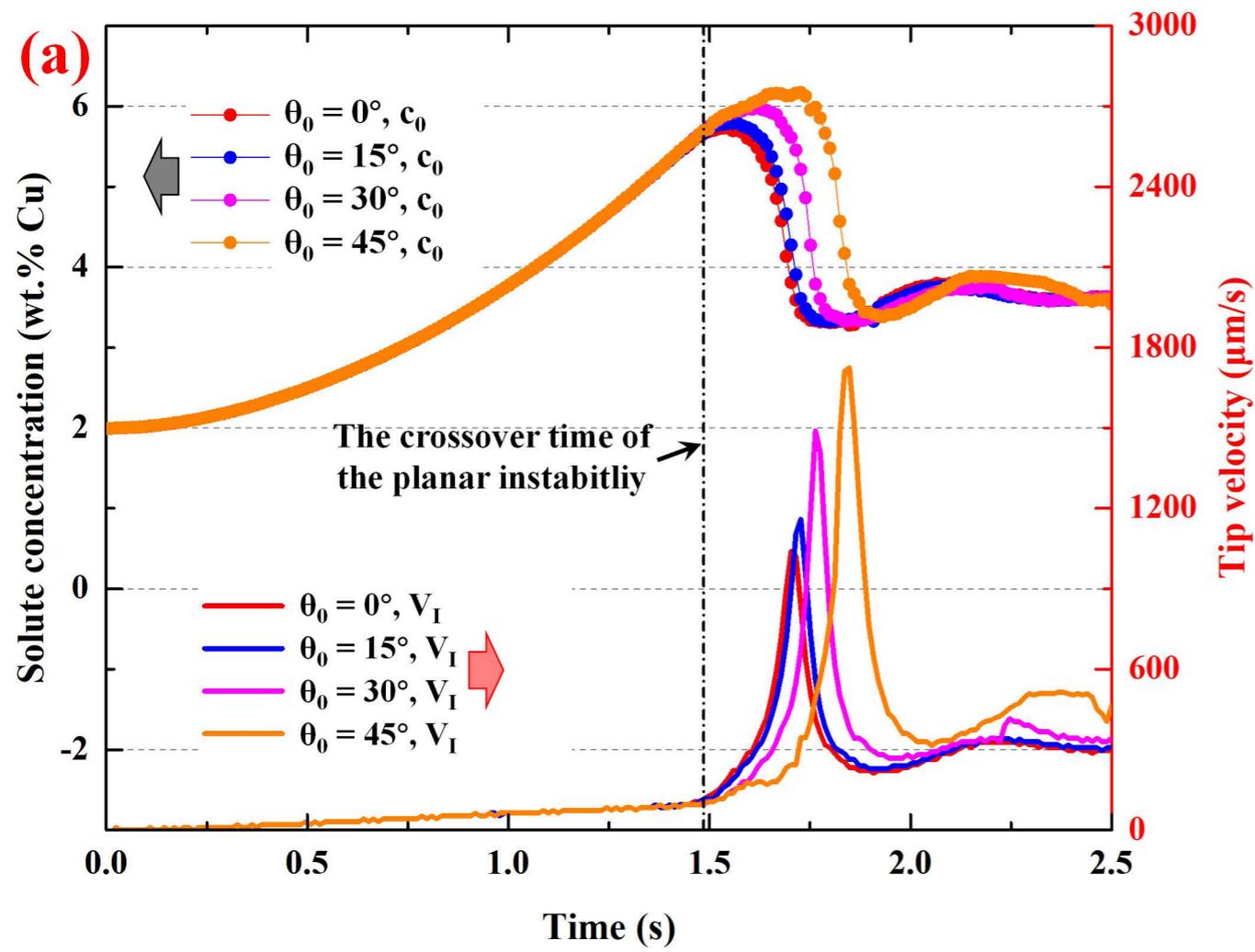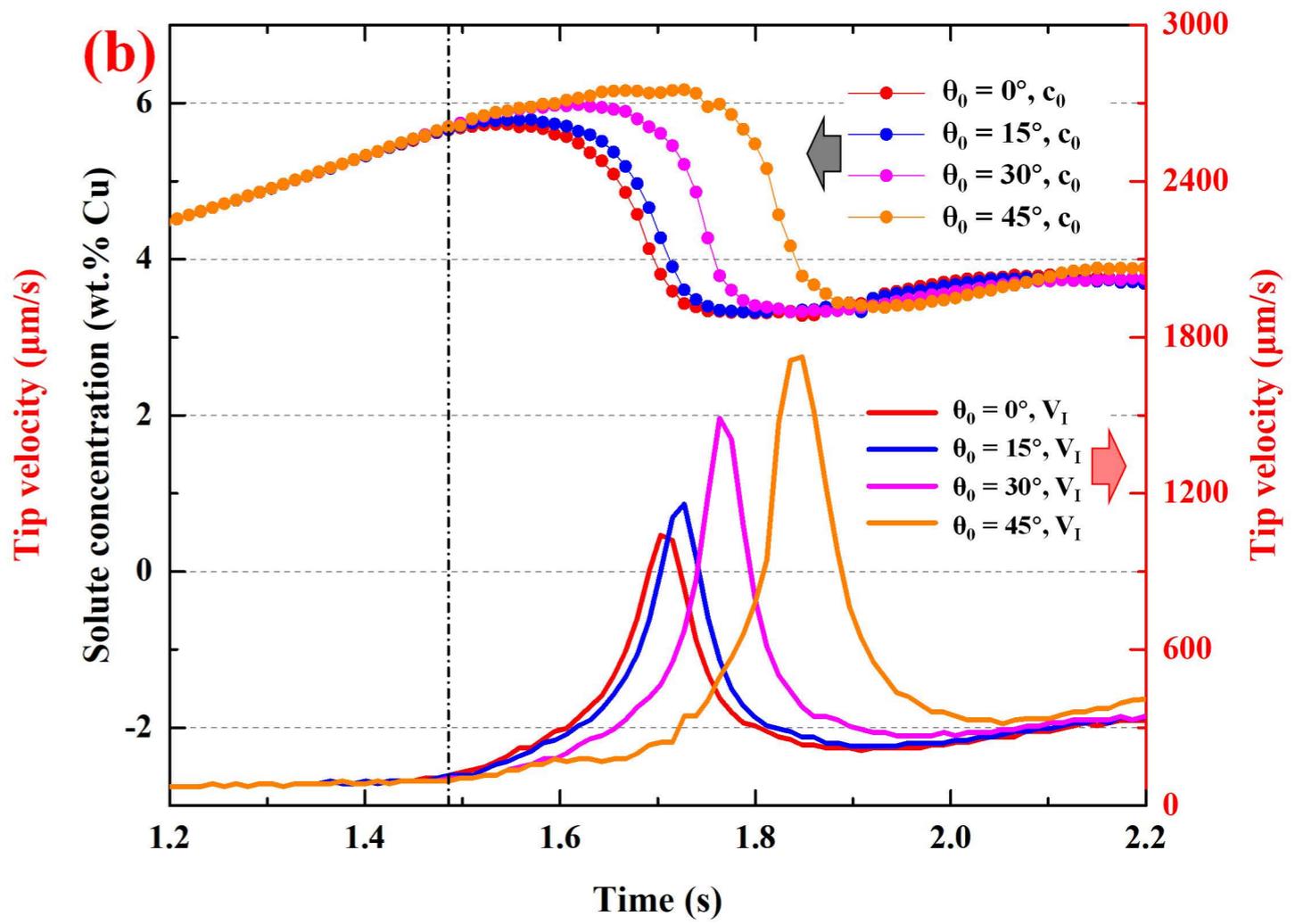

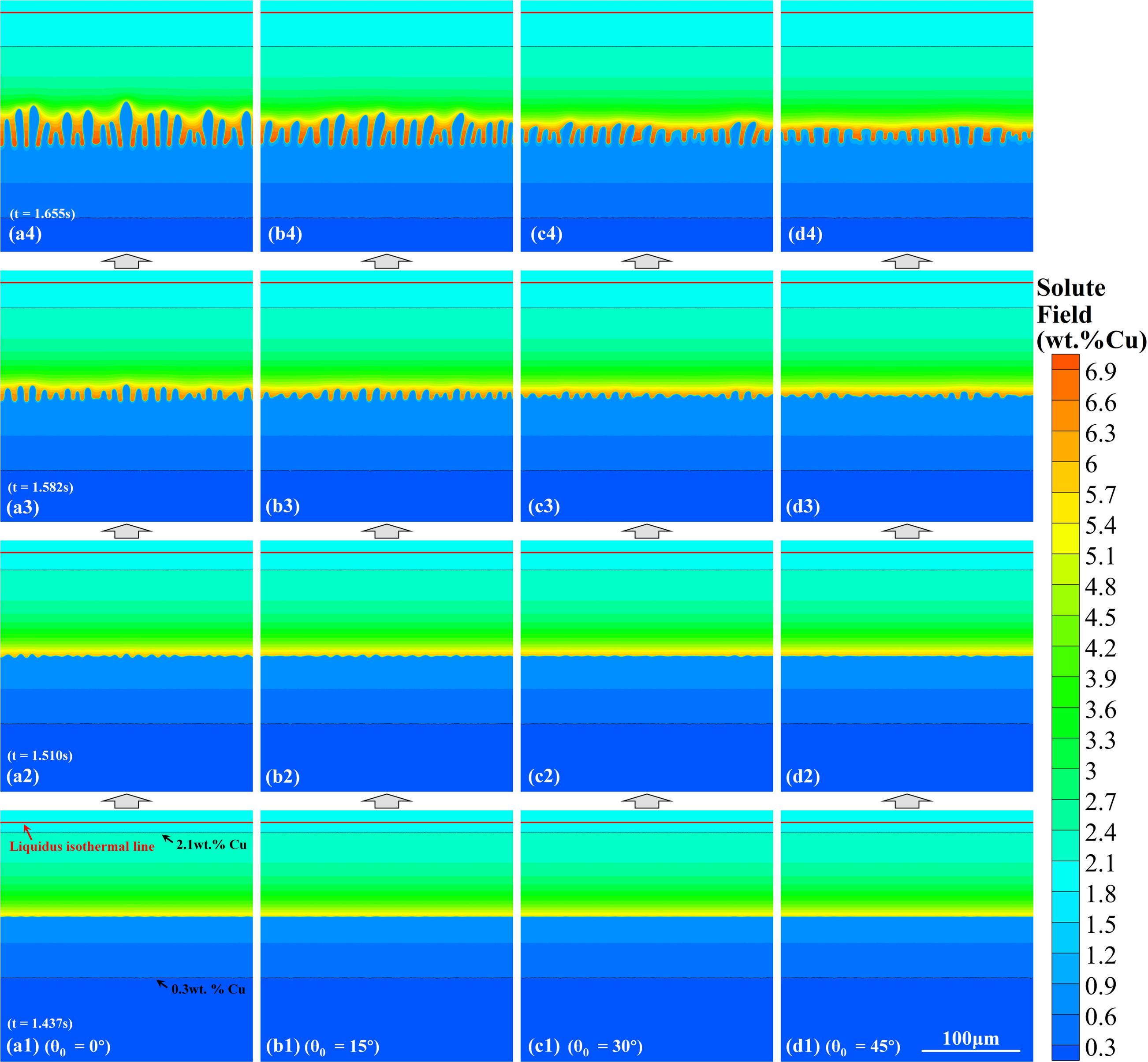